\documentclass[draft]{aastex}
\slugcomment{}

\lefthead{E. Miyata et al.}
\righthead{On the Nature of AX J2049.6+2939 and AX J2050.0+2914}
\newcommand{\Msun}{\mbox{$M_{\odot}$}}
\newcommand{\gtsim}{\,\mbox{\raisebox{0.3ex}{$>$}\hspace{-0.8em}\raisebox{-0.7ex}{$\sim$}}\,}
\newcommand{\ltsim}{\,\mbox{\raisebox{0.3ex}{$<$}\hspace{-0.8em}\raisebox{-0.7ex}{$\sim$}}\,}

\begin{document}

\title{On the Nature of AX J2049.6+2939/AX J2050.0+2914}
\author{Emi Miyata\altaffilmark{1,8},
Kouji Ohta\altaffilmark{2},
Ken'ichi Torii\altaffilmark{3},
Toshiaki Takeshima\altaffilmark{4,5},
Hiroshi Tsunemi\altaffilmark{1,8},
Takashi Hasegawa\altaffilmark{6},
and
Yasuhiro Hashimoto\altaffilmark{7}
}

\altaffiltext{1}{Department of Earth and Space Science,
        Graduate School of Science, Osaka University,
	1-1 Machikaneyama, Toyonaka, Osaka 560-0043 JAPAN;
	miyata@ess.sci.osaka-u.ac.jp, tsunemi@ess.sci.osaka-u.ac.jp}
\altaffiltext{2}{Department of Astronomy, Faculty of Science, Kyoto
        University, Sakyo-ku, Kyoto 606-8502 JAPAN; ohta@kusastro.kyoto-u.ac.jp}
\altaffiltext{3}{NASDA TKSC SURP, 2-1-1 Sengen, Tsukuba, Ibaraki
        305-8505  JAPAN; Torii.Kenichi@nasda.go.jp}
\altaffiltext{4}{Code 662, NASA Goddard Space Flight Center, Greenbelt,
        MD 20711, USA; takeshim@ginpo.gsfc.nasa.gov}
\altaffiltext{5}{Universities Space Research Association,
        7501 Forbes Blvd., Suite 206, Seabrook, MD  20706, USA}
\altaffiltext{6}{Gunma Atronomical Observatory, Ohtomo, Maebashi, Gunma
        JAPAN; hasegawa@watarase.astron.pref.gunma.jp}
\altaffiltext{7}{Carnegie Observatories, 813 Santa Barbara Street,
Pasadena, CA 91101; hashimot@vorpal.ociw.edu}
\altaffiltext{8}{CREST, Japan Science and Technology
        Corporation (JST), 2-1-6 Sengen, Tsukuba, Ibaraki 305-0047  JAPAN}
\begin{abstract}

 AX J2049.6+2939 is a compact X-ray source in the vicinity of the
 southern blow-up region of the Cygnus Loop supernova remnant (Miyata et
 al. 1998a). This source was the brightest X-ray source inside the
 Cygnus Loop observed during the ASCA survey project. The X-ray spectrum
 was well fitted by a power-law function with a photon index of $-2.1
 \pm 0.1$. Short-term timing analysis was performed and no coherent
 pulsation was found. Follow-up observations with ASCA have revealed a
 large variation in X-ray intensity by a factor of $\simeq$ 50, whereas
 the spectral shape did not change within the statistical
 uncertainties. In the second ASCA observation, we found another X-ray
 source, AX J2050.0+2941, at the north east of AX
 J2049.6+2939. During the three ASCA observations, the X-ray intensity
 of AX J2050.0+2941 varied by a factor of $\simeq$4. No coherent
 pulsations could be found for AX J2050.0+2941.

 We have performed optical photometric and spectroscopic observations in
 the vicinity of AX J2049.6+2939 at the Kitt Peak National Observatory
 (KPNO). As a result, all objects brighter than $B$-band magnitude of 22
 in the error box can be identified with normal stars. Combined with the
 X-ray results and the fact that there are no radio counterparts, AX
 J2049.6+2939 is not likely to be either an ordinary rotation-powered
 pulsar or an AGN. The nature of AX J2049.6+2939 is still unclear and
 further observations over a wide energy band are strongly required.

 As to AX J2050.0+2941, the long-term X-ray variability and the radio
 counterpart suggests that it is an AGN.

\end{abstract}


\section{Introduction}

Supernovae (SN) with a massive progenitor star leave behind a stellar
remnant. A massive star with a mass ($M$) of 10 \Msun $< M < 40$ \Msun\
leaves a neutron star while that with a M $>$ 40 \Msun\ leaves a
blackhole.  The Cygnus Loop is a well evolved supernova remnant (SNR)
and has been widely observed from radio to X-ray wavelengths. Miyata et
al. (1998b) investigated the central portion of the Cygnus Loop and
concluded that there were Si, S, and Fe rich plasmas in the center of
the Loop.  Comparing their abundance ratios with the theoretical
calculations of nucleosynthesis by a SN explosion, they estimated the
progenitor mass of $\sim$25 \Msun.  Furthermore, Miyata et al. (1999)
reported a progenitor mass of $\sim$15 \Msun\ based on the radial
distribution of heavy elements from the north-eastern region toward the
central portion. In both cases, we expect that a neutron star must have
been left behind the SN which produced the Cygnus Loop. As mentioned in
Miyata et al. (1998a; hereafter paper I), there have been several
observations to search for a neutron star associated with the Cygnus
Loop.  So far, no compact source has been established.

The ASCA Observatory is a powerful tool to discover embedded neutron
stars because of its high detection efficiency at hard X-ray region and
has discovered more than a dozen neutron stars. We have reported the
discovery of a compact X-ray source, AX J2049.6+2939, during the ASCA
survey project of the Cygnus Loop (Paper I). Its X-ray properties
suggest that it is a neutron star or an active galactic nucleus
(AGN). If it is a neutron star, it is definite that the Cygnus Loop must
have originated from a massive progenitor star.

In this paper, we performed optical photometric and spectroscopic
observations to investigate whether or not it is an AGN. We also
performed the timing analysis for follow-up ASCA observations and RXTE
observations to investigate the possibility of a neutron star which
might associate with the Cygnus Loop.

\section{Optical Observations}

Because of the possibility that the X-ray source is an AGN, we examined
it via optical observations.  We performed optical imaging observations
of the field containing AX J2049.6+2939 in the $B$-band and the $R$-band
with the WIYN 3.5m at KPNO in 1997 November.  The exposure time was 600
s for each band.  The pixel scale was $\simeq$0.$^{\!\!\prime\prime}$2.
The data were reduced and analyzed in the usual manner with {\it IRAF}.
Photometric calibration was made using Landolt's standard stars (SA 112
and SA 92) (Landolt, 1992) and a photometric error was estimated to be
0.03 mag. The magnitudes of the objects are $R \sim 17 - 20$ mag and $B
\sim 19-21$ mag with an accuracy of $\sim 0.2$ mag.  Astrometry was made
using the {\it USNO-A1} catalog and the positional error was found to be
less than 1$^{\prime\prime}$.  The $R$-band image centered on the X-ray
position is shown in figure~\ref{fig:optical} with an error circle of a
radius of 10$^{\prime\prime}$ obtained by the ROSAT HRI (paper I).

If the X-ray source is an AGN, the optical magnitude estimated from the
X-ray flux obtained in the first ASCA observation is $B = 17 -18$ mag by
a consideration of the galactic HI column density (paper I).  Thus we
selected possible optical counterparts as the objects which have a
$B$-band magnitude of $17-22$ mag located within the error circle; we
extended the possible optical magnitude range to be more secure in the
identification.  The number of candidates thus selected is four, of
which coordinates and magnitudes are listed in table 1.  We also added
one more object (No. 5) as a candidate for the spectroscopic
observation, because it is located very close to the error circle as
shown in figure~\ref{fig:optical}.  Although we searched for an object
behind the bright source in the error circle, no significant object was
found.  It should be noted that the brightest optical object ($V = 12.6$
mag) in the error circle was identified with a G-type star based on an
optical spectroscopy reported in paper I and was excluded from the
candidates in this observation.

Optical spectroscopy of the candidates was conducted in 1998 October
with the Gold Camera Spectrograph attached to the KPNO 2.1m telescope.
A grating of 158 lines mm$^{-1}$ and an order cut filter of GG400 were
used, which gave a spectral coverage of $5000 - 10000$\AA\ with a
spectral resolution of $\sim12$\AA.  One pixel covered 4.6\AA\ and
0.$^{\!\!\prime\prime}$78.  The slit width used was 2$^{\prime\prime}$
and seeing during the observing run was $\sim 2^{\prime\prime}$.  An
exposure time was 20 to 30 min and typically two or three exposures were
taken for each object.  We used an internal quartz lamp taken just after
each exposure as a flat.  Data were reduced and analyzed in the usual
manner using {\it IRAF}.  The obtained spectrum of No. 1 shows a TiO
band feature at around 6160 \AA\ as well as Mg b absorption line and is
identified with an M-type star. The spectra of Nos. 2, 3, 4, and 5 show
an Mg b absorption and a Na D absorption at $z =0$, and these objects
are identified with stars (presumably G or K stars).  The spectral types
of identified stars are also consistent with obtained $B-R$ colors.
We therefore conclude that there are no AGNs in the expected magnitude
range within the error circle.

\section{X-ray Observations}

\subsection{ASCA observations}\label{sec:asca}

We performed two follow-up observations with ASCA. (For a detailed
description of ASCA, see Tanaka, Inoue, \& Holt, 1994). In the following
ASCA analysis, we focused on the GIS data since the GIS possesses higher
detection efficiency in the hard X-ray band and a high timing resolution
below 10 keV (Ohashi et al. 1996; Makishima et al. 1996).  The second
observation with ASCA was performed on 1998 May 7. We excluded all the
data taken at elevation angle below 5$^\circ$ from the night earth rim
and 25$^\circ$ from the day earth rim, a geomagnetic cutoff rigidity
lower than 6 GV, and the region of the South Atlantic Anomaly. We also
applied a ``flare-cut'' to maximize the signal-to-noise ratio, as
described in Ishisaki et al. (Ishisaki Y. et al, 1997, in the ASCA News
letter, No.5 ).  The exposure time is $\simeq$ 25.4 ksec after the data
screening.  We constructed three kinds of images for the GIS data set: a
total count image, a non-X-ray background image, and a cosmic X-ray
background image. The non-X-ray background image was produced with the
H02-sorting method in {\sl DISPLAY45} (the detailed description of this
method is in Ishisaki (1996)). The image of the cosmic X-ray background
was extracted from the Large Sky Survey (LSS) data (Ueda et al. 1999)
acquired during the ASCA PV phase. We also extracted the non-X-ray
background image from the LSS data to produce an image of the mean
cosmic X-ray background only. In the energy band below 2 keV, the
thermal emission from the Cygnus Loop is dominant.  Thus, we extracted
background-subtracted GIS images in the energy band of $2-10$ keV shown
in figure \ref{fig:second}.  The X-ray flux of AX J2049.6+2939 is much
lower than that of the first observation (paper I).  The count rate of
AX J2049.6+2939 is $\simeq (3.9 \pm 1.3) \times 10^{-3}$ c s$^{-1}$/GIS
and is a factor of $\simeq$ 50 lower than that of the first observation.
As clearly seen in figure~\ref{fig:second}, a bright compact source can
be found at the north-eastern direction of AX J2049.6+2939. This source
was detected in the first observation but not detected with the ROSAT
HRI (see figure 1c in paper I). The location (J2000) of this source is
${\rm \alpha = 20^h 50^m 0^s}$ and ${\rm \delta = 29^\circ 41^\prime}$
with an error radius of 70$^{\prime\prime}$.  We tentatively call this
source as AX J2050.0+2941. Unfortunately, this source is outside of the
SIS FOV.

We observed the region in the vicinity of AX J2049.6+2939 again for the
third observation.  It was performed in 1998 November 10. We screened the
data using the same method as for the second observation, giving a 
total observation time of 7.7
ksec. The image obtained with the GIS is shown in figure
\ref{fig:third}. The GIS count rate of AX J2049.6+2939 is $ (2.4 \pm
2.0) \times 10^{-3}$ c s$^{-1}$/GIS and is consistent with that of the
second observation. \\

We should note that AX J2049.6+2939 was detected at only a two sigma
significance level in the SIS data.

\subsubsection{X-Ray Spectra of AX J2049.6+2939}

We extracted the X-ray spectrum of AX J2049.6+2939 and the background
spectrum from the same region in the same manner as for the first observation
(described in paper I).  We made spectra both from the GIS2 and GIS3 data and
added them together. The response matrix was created from a
count-weighted sum of GIS2 and GIS3 responses. The cross-calibration
uncertainties are less than 5 \% as may been found at the {\tt ASCA GOF web page}.
We applied several simple models to our data set and the results are described
in table 3. $N_{\rm H}$ is fixed to $3.1 \times 10^{21} {\rm cm}^{-2}$
which is the same value as that obtained in the first observation (paper
I). We should note that the fitting results were not affected when we chose
the other values that we obtained in table 1 of paper I. In all models listed in
table 3, we obtained consistent results with those of the first
observation within statistical uncertainties. Therefore, the spectral
shape seems to be unchanged whereas the flux was decreased by a factor
of $\simeq$ 50.

\subsubsection{X-Ray Spectra of AX J2050.0+2941}

We next extracted the X-ray spectrum of AX J2050.0+2941 for all three
data sets.  As shown in fig 1 (b) in paper I, AX J2050.0+2941 is
severely contaminated by AX J2049.6+2939.  The point spread function
(PSF) of the ASCA XRT has a sharp core with a broad wing structure
(Serlemitsos et al. 1995). Although there are complex structures in the PSF,
the on-axis PSF shows a well-defined axially symmetric shape. 
The 7$^\prime$ off-axis PSF shows a small
elongation along the tangential direction but is still axially 
symmetric,  as shown in figure 5b of Serlemitsos et al. (1995).  We
observed AX J2049.6+2939 and AX J2050.0+2941 within 7$^\prime$ from the
on-axis for all three observations.  Therefore, we select the region
which is on the opposite side of AX J2050.0+2941 across the location of
AX J2049.6+2939 to accumulate the background spectrum. The count rates
of AX J2050.0+2941 are $(1.3 \pm 0.2) \times 10^{-2}$ c s$^{-1}$ /GIS,
$(9 \pm 2) \times 10^{-3}$ c s$^{-1}$ /GIS, and $(3 \pm 2) \times
10^{-3}$ c s$^{-1}$ /GIS for the first, second, and third observations,
respectively.  We applied the same models as those for AX J2049.6+2939
and the results are given in table 4.  In the third observation, we could not
obtain meaningful results due to low X-ray intensity and relatively
short exposure time.  Only bremsstrahlung and power-law models are
acceptable for both data sets. We found no difference in their spectral
shapes within the statistical uncertainties. In the case of the
power-law model, the X-ray flux in the energy range of $2-10$ keV is
$(3.4\pm 0.6)\times 10^{-13}$ and $(3.9\pm 0.6)\times 10^{-13} {\rm\ erg
\ s ^{-1}\ cm^{-2}} $ for the first and the second data sets,
respectively. In the third observation, we calculated the X-ray flux
assuming the power-law model with a photon index of $-2.25$ and $N_{\rm
H}$ of $1.0 \times 10^{22} {\rm cm}^{-2}$ to be $(9 \pm 4)\times
10^{-14} {\rm\ erg \ s^{-1}\ cm^{-2}}$. AX J2050.0+2941 shows the X-ray
intensity variation by a factor of 4.

We obtained the Einstein IPC data set in the vicinity of AX J2050.0+2941
from the HEASARC/GSFC Online Service using the observation ID of 3782.
Unfortunately, AX J2050.0+2941 is located just on the support grid of
IPC and it is difficult to estimate the X-ray flux with a high degree of
accuracy. The ROSAT HRI also observed the region of AX J2050.0+2941 as
the sequence number of 500462. However, it was not detected above the 5
$\sigma$ level. This might be consistent with the fact that the obtained
$N_{\rm H}$ value was high. Thus, we could not investigate the
long--term X-ray intensity variation for AX J2050.0+2941.

\subsubsection{Short--Term Variation}

We searched for short--term intensity variations of the two compact
sources in the second observation using the GIS data.  Since we assigned
the timing bit of the GIS to be 10 for the ASCA second and third
observations, the timing resolutions for high and medium bit rate are 61
and 488 $\mu$s (Ohashi et al. 1996).  We used the data obtained both
with the high and the medium bit rate for the temporal analysis. After
applying the barycentric correction on photon arrival times, we
performed the FFT analysis with a Nyquist frequency of 1024 Hz.  We
found no coherent pulsations for neither AX J2049.6+2939 nor AX
J2050.0+2941. The 99 \% upper limits on the pulsed fraction for the
sinusoidal pulse shape are 42 \% and 36 \% for $0.7-10$ keV band and 62
\% and 55 \% for $2-10$ keV band, respectively, in the second
observation.

Because the Fourier transform method is not sensitive enough for
non-sinusoidal pulse shapes, we also tried the epoch folding search
method. Arrival times were folded into 10 phase bins for assumed trial
periods.  The epoch folding method is extremely time consuming on the
computer and we restricted our search to the period range of $5\, {\rm
s}<P<1000\, {\rm s}$. No significant pulsation was detected. Upper
limits of pulse amplitude (99 \% confidence) were obtained as 23 \%
(0.7$-$10 keV) and 44 \% (2$-$10 keV) for AX J2049.6+2939 and 21 \%
(0.7$-$10 keV) and 41 \% (2$-$10 keV) for AX J2050.0+2941, respectively.

To further examine the aperiodic variations, we extracted light curves
of 0.7$-$10 keV with the bin sizes of 512, 1024, 2048, and 4096~s for the
two sources. No significant aperiodic variations were found at the 99 \%
confidence level. In these timescales, the rms noise was found
to be 26 \% and 20 \% for AX J2049.6+2939 and AX J2050.0+2941,
respectively. These results show that there is no excess variation above
the Poisson error.

Due to poor statistics, we could not obtain a meaningful result for the
third data set.

\subsection{RXTE observation}

AX J2049.6+2939 was observed with RXTE on 1997 December 22-23.  RXTE
carries three scientific instruments: the Proportional Counter Array
(PCA), the High-Energy X-Ray Timing Experiment (HEXTE), and the All-Sky
Monitor (ASM). Detailed description of RXTE can be found in Bradt,
Rothschild, and Swank (1993).  Here we analyze the data from the PCA
which consists of five collimated (1$^\circ$ FWHM) proportional counter
units (PCUs) that contain three multianode detector layers with a
mixture of xenon and methane gas (Jahoda et al. 1996).

To maximize the signal-to-noise ratio between 2 and 10 keV ($5-27$ PH
channel), we accumulated photon events from the top xenon/methane layer
of the PCA.  We excluded all the data taken at elevation angle below
10$^\circ$ from the earth rim, the region from the South Atlantic
Anomaly (SAA) to within 30 min after the end of the SAA passage, and the
period when there was severe electron contamination ({\tt ELECTRON0,
ELECTRON1, ELECTRON2} $\ge$ 0.1). Since PCU-3 was switched off for some
part of our observation, we ignored PCU-3 data to maximize the
observation time.  This yields a total integration time of $\simeq$ 26
ks.

We constructed light curves for the $2-10$ keV band with 64, 128, and
512 s binning and estimated the background for each time bin using {\tt
pcabackest} with a background model of {\tt
pca\_bkgd\_faintl7\_e3v19990824.mdl} and {\tt
pca\_bkgd\_faint240\_e3v19990909.mdl} retrieved from the GSFC ftp site.
There is no variability in the X-ray flux during the observation. The
rms noise is 69, 64, and 64 \% for the 64, 128, and 512s binned
data. These results show that there is no excess variation above the
Poisson error.  The mean count rate is $0.53 \pm 0.06$ c s$^{-1}$ /4PCU.

We estimated the contamination from the shell emission of the Cygnus
Loop. We extracted the ASCA 2$-$5 keV image with the same method as
Miyata et al. (2000), removed the region with the radius of 4$^\prime$
centered on AX J2049.6+2939, and estimated the intensity of shell
emission there based on the region shown in Fig 1-a in paper I. We
obtained the collimator response file for the RXTE from the calibration
database (CALDB) supported by GSFC ({\tt p0coll\_96jun05.fits}) and
convolved it with the processed ASCA image. Thus, we estimated the RXTE
countrate of the shell emission of the Cygnus Loop to be $\sim 0.3 - 1
\times 10^{-3}$ c s$^{-1}$ / 4PCU assuming the plasma temperature ranges
from 0.3$-$0.8 keV (Miyata 1996). Therefore, the contamination of
thermal emission can be negligibly small.

As mentioned in section~\ref{sec:asca}, we found a hard X-ray compact
source, AX J2050.0+2941, near AX J2049.6+2939. AX J2050.0+2941 shows
X-ray variability by a factor of 4.  Since the angular separation is
less than 7$^\prime$, both sources were simultaneously observed with
RXTE. Assuming the X-ray flux of AX J2050.0+2941 to decrease linearly
from the first to the second observation, the X-ray flux of AX
J2050.0+2941 at the RXTE observation phase can be estimated to be $1.1
\times 10^{-12} {\rm\ erg \ s ^{-1}\ cm^{-2}} $. We assumed the X-ray
spectrum of AX J2049.6+2939 to be a power-law spectrum with a photon
index of $-2.1$ and $N_{\rm H}$ to be $3.1\times 10^{21} {\rm cm^{-2}}$
(paper I), resulting in the X-ray flux of AX J2049.6+2939 to be $(3 \pm
2)\times 10^{-13}$ erg ${\rm cm^{-2} \ s^{-1}}$.  We should note that we
take into account the 20 \% cross calibration uncertainties between ASCA
and RXTE.  This value is an order of magnitude lower than that of the
ASCA first observation and is similar to that of the second ASCA
observation.

We have performed a FFT analysis for the frequency range of
$10^{-3}-$500 Hz on the barycenter corrected time series data of 2$-$10
keV detected in the top xenon/methane layer.  No significant coherent
pulsation was found.  The poor signal to noise statistics due to the
relatively low X-ray flux and high background rate ($\gtsim$ 95 \% of
the total counts) did not allow us to constrain the better upper limit
of the coherent pulsations than those with the first and second ASCA
observations.

We also tried the epoch folding search method for the RXTE data in the same way
as that for ASCA. Although we searched for the period range $5{\rm
s}<P<1000{\rm s}$, the $\chi^2$ values were found to become large for the
period range $P>20$~s due to the time variable background
counts. We therefore restricted the search range to 
$5\, {\rm s}<P<20\,{\rm s}$. An upper limit of pulse amplitude was found to
be $\sim$ 70 \% of the estimated source count rate of AX~J2049.6+2939.

\subsection{Long--Term X-ray Variation of AX J2049.6+2939}

We constructed the long--term light curve of AX J2049.6+2939, using the
Einstein IPC, the ROSAT HRI, RXTE PCA, and ASCA GIS data, which span a
period of 19 years. For the IPC and HRI observations, we extracted the
count rate of AX J2049.6+2939 while the background count rate was
estimated from regions where the intensity of thermal emission from the
Cygnus Loop was similar to that of AX J2049.6+2939.  We assumed a
power-law function and took into account the error regions obtained in
paper I to calculate the X-ray flux.  Figure~\ref{fig:long-term} shows
the obtained long--term light curve of AX J2049.6+2939. We find that the
X-ray flux of AX J2049.6+2939 generally decreases during 19 years while
a ``flare event'' occurred during the first ASCA observation.

We should note that the X-ray variability shown in
figure~\ref{fig:long-term} contains a lot of uncertainties: insufficient
(or no) cross calibration, different energy bands, assumed model and
$N_{\rm H}$. However, we can still confirm a large X-ray intensity
variation with the ASCA data, solely.

\section{Discussion}

% Miyata et al. (1998b) estimated a progenitor mass of the Cygnus Loop to
% be 25 \Msun\ based on the analysis of the enhanced abundance at the
% center portion. Miyata, Tsunemi (1999) suggested a progenitor mass of 15
% \Msun\ based on the enhanced abundance at the shell region. We therefore
% conclude that the Cygnus Loop must have originated from a type II SN
% which, we expect, should have left behind a stellar remnant.

\subsection{Identification of AX J2050.0+2941}

AX J2050.0+2941 shows an X-ray intensity variation of a factor of 4
during the ASCA observations.  It shows no short--term variability.  The
X-ray spectrum of AX J2050.0+2941 can be fitted by a power-law function
with a photon index of $\simeq -2.2$ or a thermal bremsstrahlung with
$kT_{\rm e}$ of 3 keV.

As mentioned in paper I, radio emission was detected in the vicinity
of AX J2050.0+2941 using {\it Skyview} supported by HEASARC/GSFC.
Williams et al. (1992) analyzed both radio-loud and radio-quiet quasars
with Ginga in the energy range of $2-20$ keV and found a correlation
between the radio-loudness and the X-ray spectral index.  The X-ray
spectra of radio-loud and radio-quiet quasars were well represented with
a power-law function with an index of $\sim -2.0$ and $\sim -1.6$,
respectively.  Since we have found a radio counterpart of AX
J2050.0+2941, it is likely to be a radio-loud quasar.  If AX
J2050.0+2941 is an AGN, the reddening-corrected optical magnitude
estimated from the X-ray flux is $17 \ltsim B \ltsim 21$ mag (Zamorani
et al. 1981). We then checked for a counterpart in optical wavelengths
using the {\it USNO-A1} catalog. However, AX J2050.0+2941 is detected
only in the ASCA observation and the error circle is relatively large
(70$^{\prime\prime}$).  More than forty objects were found within the
error circle. We therefore need further X-ray observations of AX
J2050.0+2941 with Chandra or XMM-Newton to determine the position of AX
J2050.0+2941 with high accuracy for its identification.

\subsection{Identification of AX J2049.6+2939}

We have found a long-term X-ray variability of over 19 years for AX
J2049.6+2939.  Only in the sequence of ASCA observations, AX
J2049.6+2939 shows a long-term X-ray variability with a factor of
$\simeq$ 50 but no short-term X-ray variability. The X-ray spectrum can
be fitted by a power-law function with an index of $\simeq -2.3$ or a
thermal bremsstrahlung with $kT_{\rm e}$ of 4 keV for three
observations. Such a large X-ray variability suggests that it is not
likely to be an ordinary rotation-powered pulsar but it could be an AGN.

We have performed optical photometric and spectroscopic
observations in the vicinity of AX J2049.6+2939 and found no AGNs within
its error circle. Thus, we conclude that AX J2049.6+2939 is not likely
to be an AGN.

We should note, however, that we performed the optical photometric
observations in 1997 November where the expected X-ray flux might be
lower than that of the ASCA first observation. We can estimate the X-ray
flux of AX J2049.6+2939 at 1997 November to be $\simeq 2 \times
10^{-12}$ erg s$^{-1}$ cm$^{-2}$ from figure \ref{fig:long-term}
assuming a smooth decrease of the X-ray flux from the ASCA first
observation to the ASCA second observation. The optical magnitude
expected from this value is $B = 17 -19$ mag (Zamorani et al. 1981). We
have investigated five objects within the error circle whose $B$-band
magnitude range $17-22$ mag based on WIYN observations. So, even if the
X-ray flux is lower than our estimate, we can conclude that no AGNs are
found within the error circle.

We investigated the possibility of the G-type star considering our new
X-ray observations. The X-ray spectrum of the G-type star is well
represented by an isothermal, two-temperature, or DEM (differential
emission measure) thermal model. The $kT_{\rm e}$ values range from
$\simeq$ 80 eV to $\simeq$2.5 keV and abundances of heavy elements are
roughly solar values (G\"udel et al 1997). The $kT_{\rm e}$ value and
abundances of AX J2049.6+2939 are a little higher and much lower,
respectively, than those of G-type stars if we apply a thin thermal
model (paper I).  For these stars, flares with much larger luminosities
have been observed where $kT_{\rm e}$ and emission measure becomes
larger than those of quiescent state. However, the timescales of their
flare are roughly several tens of minutes or hours which is much
shorter than `flare event'' seen in figure \ref{fig:long-term}.
Therefore, we again conclude that AX J2049.6+2939 is not likely to be a
G-type star.

It is well known that the X-ray pulsar 1E 2259+586 in the supernova
remnant G109.1$-$1.0 shows a flux variation of a factor of a few on
timescales of a few years (Corbet et al. 1995). This pulsar belongs to a
member of anomalous X-ray pulsars (AXPs).  Members of this group have
several common properties including a relatively long pulse period of
$6-12$ s and a steep X-ray spectrum characterized by a power-law
function with a photon index of $-2.5\sim -4$. Comparing these
characteristics, long-term X-ray variability seems to be too large for
AXPs, there is no coherent pulsation, and the spectrum is flatter than
those of AXPs. Thus, AX J2049.6+2939 is not likely to be an AXP.

A remarkably similar situation is found for the compact source, 1E
161348-5055, in the supernova remnant RCW 103. Gotthelf, Petre \&
Vasisht (1999) reported the X-ray variability of 1E 161348-5055. They
found that its X-ray flux decreased by an order of magnitude over a
period of 18 years. No coherent pulsation has been detected from this
source (Gotthelf et al 1999 and references therein). These
characteristics are similar to those of AX J2049.6+2939 whereas the
X-ray spectrum of 1E 161348-5055 is much softer than that of AX
J2049.6+2939. They propose that 1E 161348-5055 is either a cooling
neutron star (but less plausible), an advection dominated accretion flow
(ADAF) around a blackhole , or an AXP.  If we apply the same analogy for
AX J2049.6+2939, this source might well be accreting through an ADAF.
Since there are CO molecular clouds at the western edge of the Cygnus
Loop (Scoville et al. 1977), such a model might be applicable to AX
J2049.6+2939. The X-ray spectra of AX J2049.6+2939 are consistent with
that expected from the ADAF model (Narayan, Barret, \& McClintock 1997;
Manmoto, Mineshige, \& Kusunose 1997). The ADAF model also predicts that
the shape of the X-ray spectrum is independent of the X-ray flux. Thus,
there is no definite evidence for ADAF but apparent X-ray features are
consistent with those expected from the ADAF model.

\section{Conclusion}

We performed wide-band spectroscopic observations in the vicinity of AX
J2049.6+2939. Based on the optical photometric and spectroscopic
observations, we can rule out the possibility of an AGN. AX J2049.6+2939
shows a large X-ray variability over 19 years, which excludes the
possibility of an ordinary rotation-powered pulsar.  Thus, the ADAF
model would be a plausible one, although no definite evidence is
obtained.  As calculated by Narayan, Barret, \& McClintock (1997) and
Manmoto, Mineshige, \& Kusunose (1997), the spectrum generated by ADAF
shows the synchrotron peak which comprises the Rayleigh-Jeans slope and
the optically thin synchrotron emission in the radio regime. The
Comptonization of the synchrotron photons leads to one or two peaks from
IR to UV regimes. In the X-ray and $\gamma$ ray regimes, the
bremsstrahlung emission plus photons suffering multiple Compton
scattering and emission from saturated Comptonized photons are dominant.
Therefore, further simultaneous observations over a wide energy band are
essential to investigate its nature.

We have found another hard X-ray emitting compact source, AX
J2050.0+2941, which is in a direction north-eastern from AX
J2049.6+2939.  We also found radio (4850MHz) emission in the vicinity of
AX J2050.0+2941.  Due to its large error circle, we found more than
forty objects for the optical candidate of AX J2050.0+2941. We therefore
need further X-ray observations in the vicinity of AX J2050.0+2941 with
Chandra or XMM-Newton in order to determine its position more
accurately.  It would be essential to investigate the relationship with
the Cygnus Loop.

\vspace{1pc}\par
We thank Profs. S. Kitamoto, F. Nagase, and S. Mineshige for invaluable
comments and suggestions.  We appreciate Mr. C. Baluta for careful
reading of manuscript. We thank {\sl ASCA\_ANL} and {\sl DISPLAY 45}
developing team members, especially Dr. Ishisaki.  Part of this research
has made use of data obtained through the High Energy Astrophysics
Science Archive Research Center Online Service, provided by the
NASA/Goddard Space Flight Center. This research is partially supported
by ACT-JST Program, Japan Science and Technology Corporation.

\clearpage

\begin{deluxetable}{ccccc}
\tabletypesize{\normalsize}
\tablecaption{Optical photometory for counterpart candidates of AX
 J2049.6+2939}
\tablewidth{0pt}
\tablehead{
\colhead{Designation} & \colhead{Right Ascension} & \colhead{Declination} &
\colhead{$B$-mag} & \colhead{$R$-mag}\\
\colhead{} & \colhead{(J2000)} & \colhead{(J2000)} & \colhead{} & \colhead{} 
}
\startdata
   No.1\dotfill & 20$^{\rm h}$ 49$^{\rm m}$ 34.$^{\!\!\rm s}$84 &
   29$^\circ$ 38$^{\prime}$ 53.$\!\!^{\prime\prime}$8 & 22.0 & 19.6 \\
   No.2\dotfill & 20$^{\rm h}$ 49$^{\rm m}$ 35.$^{\!\!\rm s}$01 &
   29$^\circ$ 38$^{\prime}$ 48.$\!\!^{\prime\prime}$7 & 21.0 & 19.2 \\
   No.3\dotfill & 20$^{\rm h}$ 49$^{\rm m}$ 35.$^{\!\!\rm s}$24 &
   29$^\circ$ 38$^{\prime}$ 41.$\!\!^{\prime\prime}$8 & 20.9 & 19.6 \\
   No.4\dotfill & 20$^{\rm h}$ 49$^{\rm m}$ 36.$^{\!\!\rm s}$02 &
   29$^\circ$ 38$^{\prime}$ 49.$\!\!^{\prime\prime}$5 & 20.9 & 19.7 \\
   No.5\dotfill & 20$^{\rm h}$ 49$^{\rm m}$ 36.$^{\!\!\rm s}$05 &
   29$^\circ$ 38$^{\prime}$ 43.$\!\!^{\prime\prime}$9 & 18.9 & 17.0 \\
 \enddata
\end{deluxetable}

\clearpage

\begin{deluxetable}{cccc}
\tabletypesize{\normalsize}
\tablecaption{ASCA Observations}
\tablewidth{0pt}
\tablehead{
\colhead{Pointing} & \colhead{Sequencer Number} & \colhead{Date} &
\colhead{GIS exposure time [ks]}
}
\startdata
   1st\dotfill & 55026080 & 1997 June 03 & 7.6 \\
   2nd\dotfill & 56036000 & 1998 May 07  & 25.7 \\
   3rd\dotfill & 56036010 & 1998 Nov 10 & 7.7 \\
 \enddata
\end{deluxetable}

\clearpage

\begin{deluxetable}{cccc}
\tabletypesize{\small}
\tablecaption{Fitting results of AX J2049.6+2939}
\tablewidth{0pt}
\tablehead{
\colhead{Model} &\multicolumn{3}{c}{AX J2049.6+2939}\\
\cline{2-4}
\colhead{} & \colhead{Parameter} & \colhead{Normalization} &
\colhead{$\chi^2$ / d.o.f}
}
\startdata
   Blackbody\dotfill & $kT_{\rm e} = 0.7 \pm 0.2$ keV 
   & $(1.1^{+0.2}_{-0.3}) \times 10^{-35}$  erg s$^{-1}\ ^{\rm a}$
   & 22.1 / 33 \\
   Bremsstrahlung\dotfill  
   & $kT_{\rm e} = 3.3^{+4.6}_{-1.5}$ keV
   & $4^{+2}_{-1} \times 10^{-2}$ cm$^{-6}$ pc
   & 19.6 / 33 \\
   Power law\dotfill & $\Gamma$ = $-2.5^{+0.5}_{-0.7}$ 
   & $(9^{+6}_{-4}) \times 10^{-5}$ photons keV$^{-1}$ cm$^{-2}\ ^{\rm b}$
   & 20.3 / 33 \\
 \enddata

\tablecomments{Quoted errors are at 90 \% confidence level.}
\tablenotetext{a}{Distance to the source is assumed to be 770 pc.}
\tablenotetext{b}{Flux is shown at 1 keV.}
\end{deluxetable}

\clearpage

\begin{deluxetable}{ccccccccc}
 \tabletypesize{\tiny}
 \tablecaption{Fitting results of AX J2050.0+2941}
 \tablewidth{0pt}
 \tablehead{
 \colhead{Model} &\multicolumn{4}{c}{First observation} &
 \multicolumn{4}{c}{Second observation}\\
 \cline{2-5} \cline{6-9}
 \colhead{}& \colhead{Parameters} & \colhead{$N_{\rm H}\ ^{\rm a}$} &
 \colhead{Normalization} & \colhead{$\chi^2$ / d.o.f} &
 \colhead{Parameters} & \colhead{$N_{\rm H}\ ^{\rm a}$} &
 \colhead{Normalization} & \colhead{$\chi^2$ / d.o.f}
 }
 \startdata
 Blackbody\dotfill 
 & $0.7 \pm 0.1$ keV & $\ltsim$4 
 &  $(2.8^{+0.7}_{-0.5}) \times 10^{-33}\ ^{\rm b}$
 & 38.0 / 27 & 
 $1.0 \pm 0.2$ keV  & $\ltsim$3
 & $(2.6 \pm 0.5) \times 10^{-35}\ ^{\rm b}$ & 54.6 / 40 \\
 Bremsstrahlung\dotfill  
 & $2^{+2}_{-1}$ keV & $5^{+5}_{-3}$
 & $15^{+16}_{-6} \ ^{\rm c}$ & 35.2 / 27 & 
 $\gtsim 4$ keV & $\ltsim$8
 & $(5.7^{+3.3}_{-0.8}) \times 10^{-2}\ ^{\rm c}$ & 47.6 / 40 \\
 Power law\dotfill 
 & $\Gamma$ = $-2.9^{+0.6}_{-0.9}$ & $9^{+7}_{-5}$
 & $4^{+7}_{-2} \times 10^{-2}\ ^{\rm d}$
 & 35.6 / 27 & 
 $\Gamma$ = $-1.6^{+0.4}_{-0.8}$ & $\ltsim$11
 & $8^{+17}_{-4} \times 10^{-5}\ ^{\rm d}$
 & 47.9 / 40 \\
 \enddata
 \tablecomments{Quoted errors are at 90 \% confidence level.}
 \tablenotetext{a}{$N_{\rm H}$ is in unit of $10^{21} \ {\rm cm}^{-2}$.}
 \tablenotetext{b}{Distance to the source is assumed to be 770 pc and
 luminosity is shown in unit of erg s$^{-1}$.}
 \tablenotetext{c}{Emission measure is shown in unit of cm$^{-6}$ pc.}
 \tablenotetext{d}{Flux at 1 keV is shown in unit of photons keV$^{-1}$
 cm$^{-2}$.}
\end{deluxetable}

\clearpage

\begin{figure}

\centerline{FIGURE CAPTIONS}

 \caption{Optical ($R$-band) finding chart for counterpart
 candidates.  North is to the top, and east is to the left.  Coordinates
 of the candidates are listed in Table 1.  Error circle for the X-ray
 source is shown with a radius of 10$^{\prime \prime}$.}
 \label{fig:optical}

 \caption{Background-subtracted GIS image in the energy band
 of $2-10$ keV in the second observation
 of AX J2049.6+2939. The image was smoothed with a Gaussian function of
 $\sigma = 1^\prime$. The location of AX J2049.6+2939 is marked with
 a white cross.}
 \label{fig:second}

 \caption{Same as figure~\ref{fig:second} but for the third observation.}
 \label{fig:third}

 \caption{Long-term X-ray variability of AX J2049.6+2939.}
 \label{fig:long-term}
\end{figure}

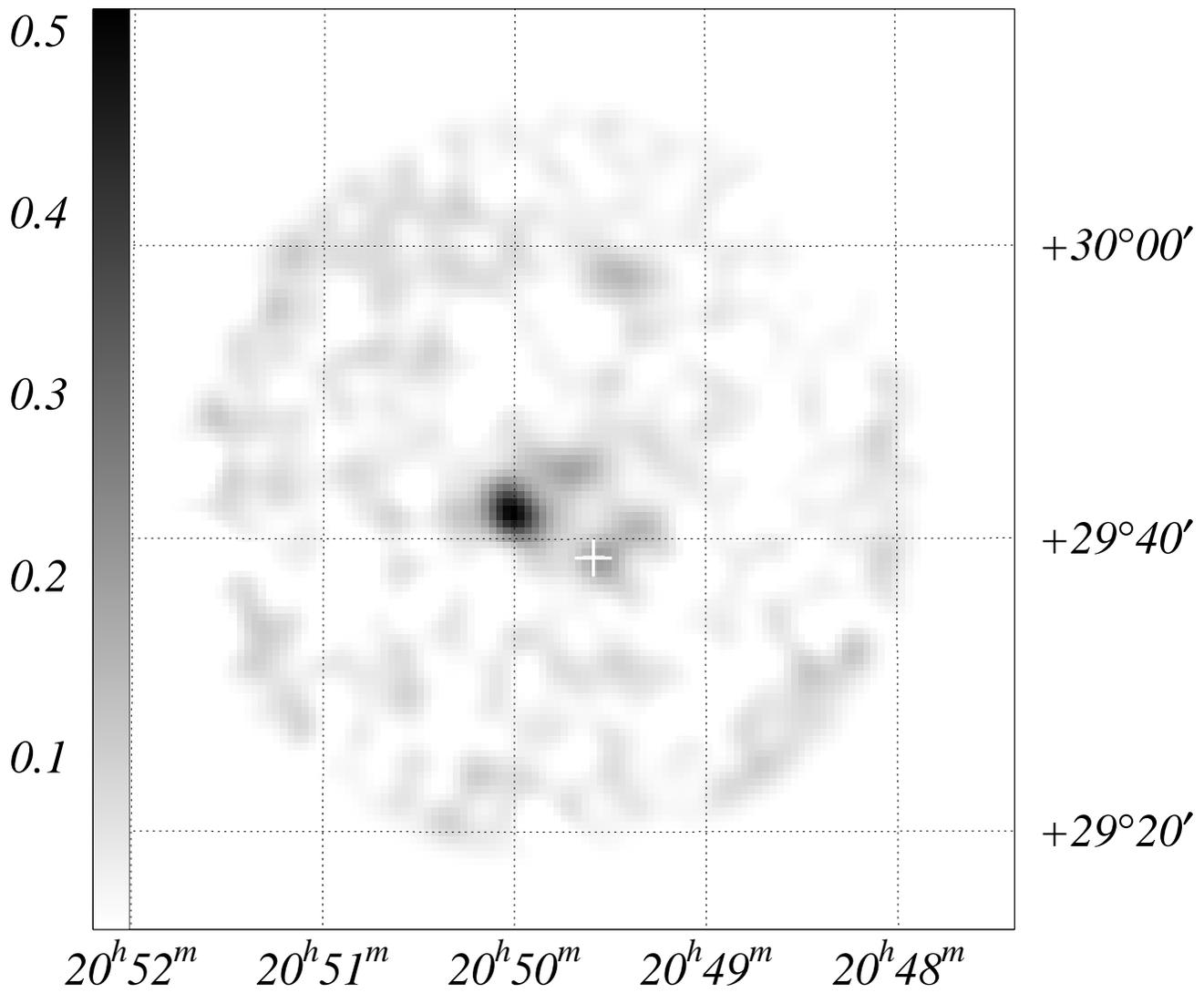
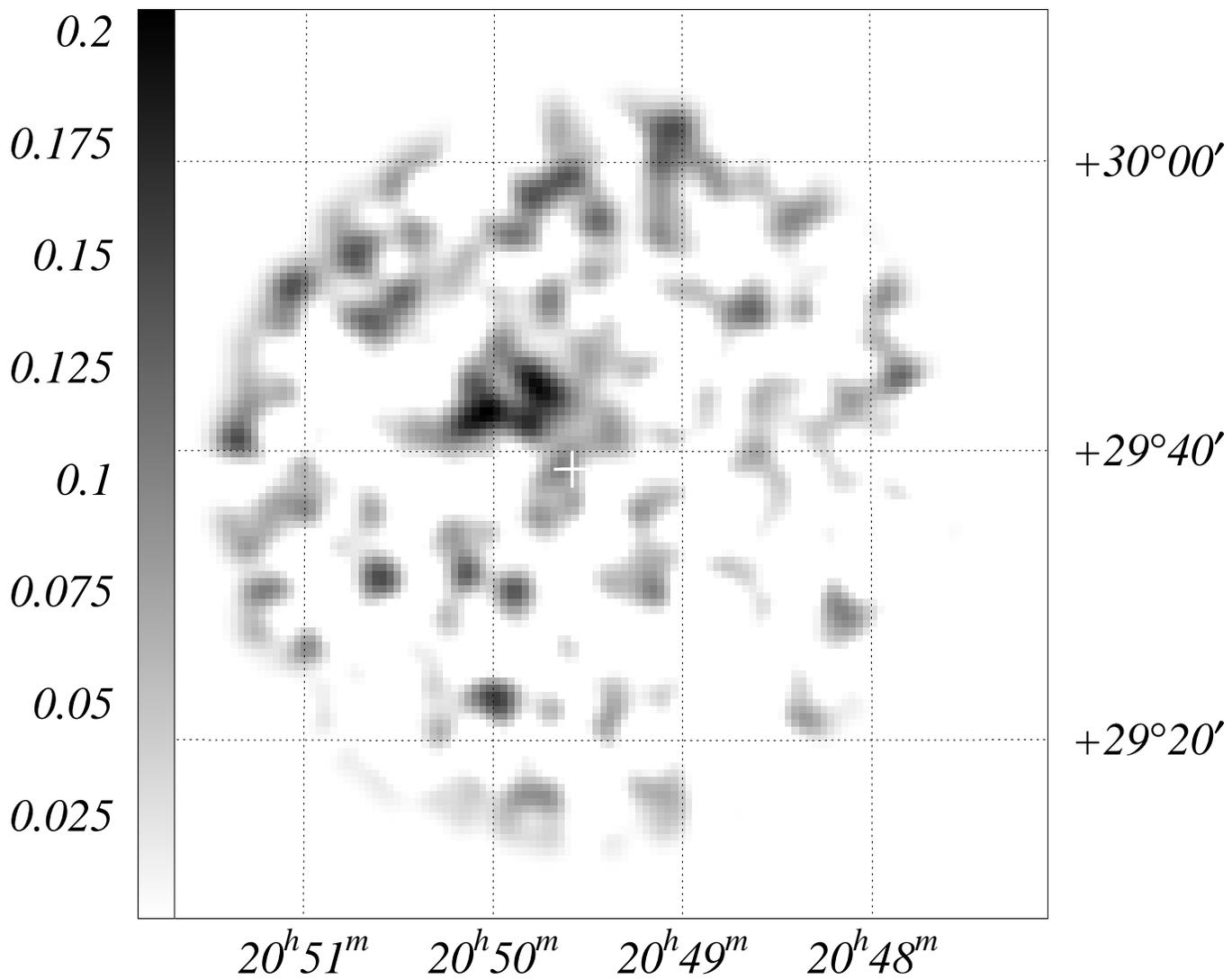
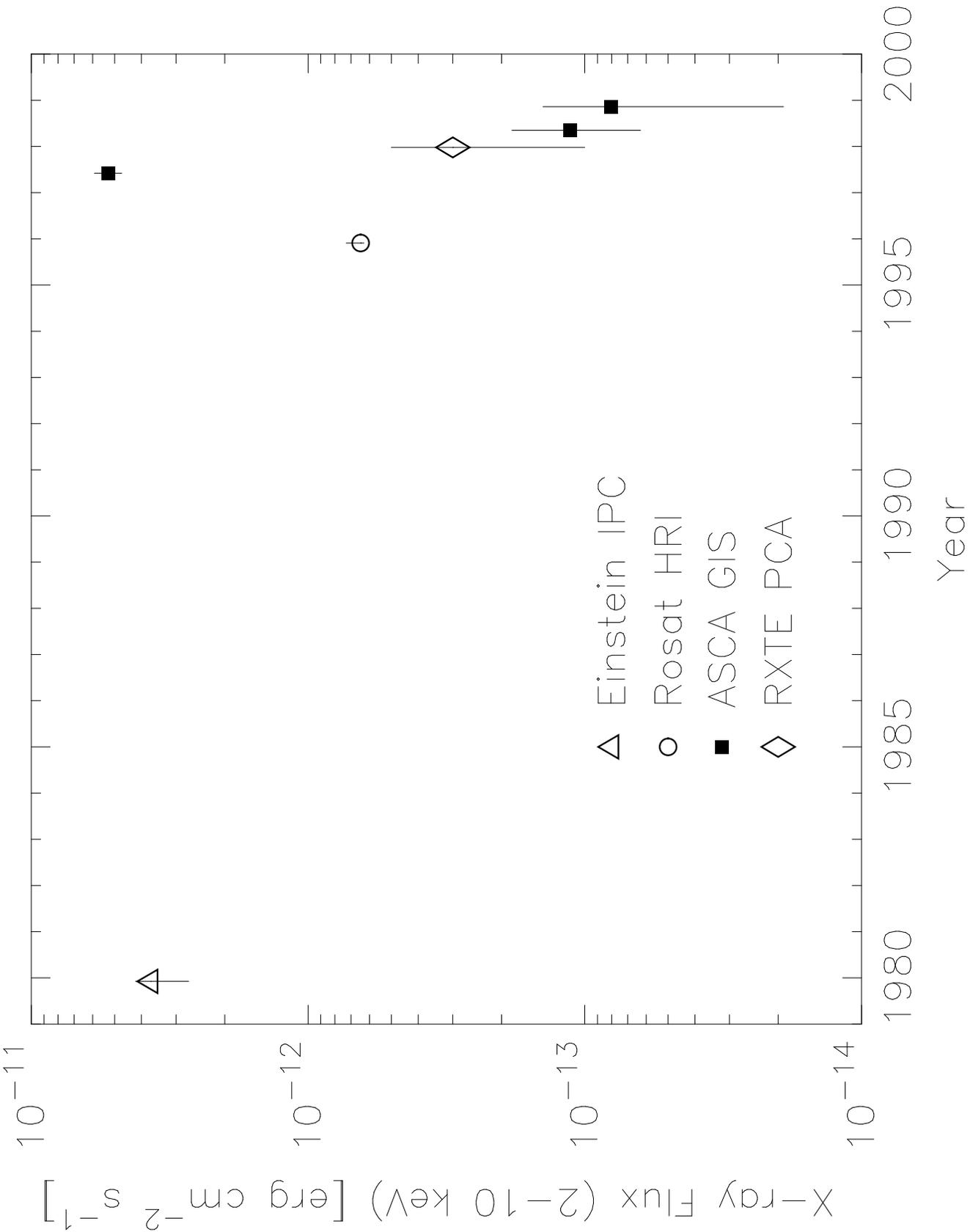
\end{document}